\documentclass{JHEP3}

\usepackage{cite}
\usepackage{epsfig}
\usepackage{amsmath}
\usepackage{array}
\usepackage{axodraw}
\usepackage{bbm}

\def\mathswitch#1{\relax\ifmmode#1\else$#1$\fi}
\def\mathswitchr#1{\relax\ifmmode{\mathrm{#1}}\else$\mathrm{#1}$\fi}

\newcommand{\tev}{\,\, \mathrm{TeV}}
\newcommand{\gev}{\,\, \mathrm{GeV}}

\newcommand{\gesim}{\gtrsim}

\title{Muon Excess in Cosmic Rays and at CDF: Signs for a Hidden Sector?}

\author{Thomas Gehrmann, Nicolas Greiner and Pedro Schwaller \\ Institut f\"ur Theoretische Physik,
        Universit\"at Z\"urich, \\ Winterthurerstrasse 190, CH-8057
        Z\"urich, Switzerland 
        }
\abstract{In this letter we discuss a certain class of Hidden Valley models
  where the dynamics in the hidden sector are close to a strongly coupled
  conformal fixed point. We show that these models can explain the excess in
  cosmic ray muon events with high muon multiplicities that has been reported
  from cosmic ray studies with the ALEPH and DELPHI detectors. 
We further point out that these models can also at least partially account for the excess in multi-muon events that was found by the CDF experiment. Finally we discuss possible signatures of these models at the LHC.}
\preprint{ZU-TH 21/08}
\begin{document}


\section{Introduction}
Recently the CDF collaboration presented a study of multi-muon events in
the $b\bar{b}$ search channel \cite{Aaltonen:2008qz}. 
They considered events with at least two central muons ($|\eta|<0.7$), 
a transverse momentum of $p_T\ge 3 \gev$,
and an invariant mass of the muon pair greater than $ 5 \gev$. Events where both 
triggered muons are produced within the beam pipe are well understood and the
measured event rate is in agreement with the theoretical prediction.

Once the requirement that the two muons are produced in the beam pipe is removed, the observed number of
events lies above the theoretical prediction - about 150000 events remain
unexplained and have been termed the \textit{ghost sample} in \cite{Aaltonen:2008qz}.
They further observed that within the ghost sample, the number of additional muons within
a $\cos \theta \geq 0.8$ cone around a primary muon is unexpectedly large, more than
four times what one would expect for an ordinary standard model event.

CDF is not the only experiment where an excess of multi
muon final states in hadronic collisions
was found and may hint to new physics. In \cite{Abdallah:2007fk} the
DELPHI collaboration at LEP studied
multi-muon events coming from cosmic ray showers. 
They compared the
number of events with a certain muon multiplicity with the prediction of
a Standard Model Monte Carlo simulation. They found that while the simulation
agrees with the number of events for low and medium muon multiplicities, the
number of events with the highest muon multiplicities, $(N_\mu \geq 45)$, clearly
shows an excess over the theoretical prediction. 
Depending on the model it is an excess of up to $3\sigma$ compared to the Standard Model. 
\\
The findings of the DELPHI study \cite{Abdallah:2007fk} nicely confirm the result
of an earlier study by the ALEPH collaboration, which observed \cite{Avati:2000mn}
an increase of high multiplicity muon events compared
to the Monte Carlo 
prediction. Although they can explain the
integral multiplicity distribution for multiplicities of $N_\mu \le 60$ they fail to
explain the highest multiplicities. 

Hidden valley models \cite{Strassler:2006im} are a class of hidden sector models with strong
dynamics and a spectrum light enough to be observable at current or future collider experiments.
Their phenomenology, in particular at hadron colliders, has been studied in \cite{Han:2007ae,Strassler:2006ri,Krolikowski:2008qa,Strassler:2008fv,SanchisLozano:2008te,Wells:2008xg}. 
In \cite{Strassler:2008jq} it has been pointed out that the CDF multi-muon excess
could be explained by a Hidden Valley model.

In this paper we analyse a certain Hidden Valley scenario with emphasis on its 
phenomenology in cosmic ray showers. We show that the excess of multi-muon
events can be explained by the production of a heavy particle that
decays into the hidden sector. We further argue that the assumptions needed to
account for the cosmic muon excess also offer the possibility that this model explains
the multi-muon excess reported by CDF.

This paper is organised as follows. After discussing the muon excess in cosmic ray showers in the next section, we show how conformal Hidden Valley models can explain that excess and argue that it could also explain the CDF anomaly. In section 4 we discuss possible LHC signals of these models before we conclude in section 5. 

\section{Cosmic Rays and Muons}
Cosmic rays originate from extraterrestrial sources that seem to be
uniformly
distributed. These particles can be electromagnetic particles such as
electrons
and photons as well as strongly interacting particles such as protons and
nuclei.
For energies above $\sim 1 \gev$ the electromagnetic component is
negligible so that
the usual assumption is that the incident particle is a proton or a
nucleus. The most
stable nucleus and the heaviest that can be produced by nuclear fusion
is iron. Therefore
one assumes that the cosmic rays consist of a mixture of nuclei from single protons
to iron. A more detailed discussion of cosmic ray showers and simulations
of the muon density profile can be found in \cite{Travnicek:2004sc}.

Detected muons from cosmic ray showers are mainly decay products from
hadrons produced
after the initial collision of a cosmic particle with a nucleus of the
atmosphere. The setup of the DELPHI experiment allows to observe
events with cosmic ray energies between $10^{14}-10^{18}$\,\,eV. 
These high energies assure that the muon bundle is strongly boosted and this
restricts the radius of the cone of the decay products.
The
hadronic showering and the subsequent decay into muons can be simulated
using Monte Carlo methods. The DELPHI collaboration used QGSJET01 
\cite{qgsjet} to simulate the initial collision and the COSIKA program\cite{corsika} 
to model the atmospheric shower.
They have varied the center of  the shower
within a certain radius around the detector. Their result was that the optimal
radius is around $R=200\,\,$m. If the radius is smaller, one looses events with
low multiplicities whereas if the radius is larger one needs larger data
samples
to also get events with high multiplicities. For that value of the radius
one looses less than $0.5\%$ of the events with the lowest multiplicity
of $N_{\mu}=4$.
This leads to the conclusion that the muons in a cosmic shower are typically 
concentrated in a radius of $200\,\,$m from the center. 
From these simulations one
can nicely
see that for very low multiplicities the data fits the model with a
proton-induced
shower whereas for multiplicities with $5\le N_{\mu}\le 50$ the data is
well described by an iron induced shower. However for multiplicities
with $N_{\mu}\ge50$
the simulation fails to describe the data.\\
A similar  characteristic behavior has been found by analysing cosmic 
ray data of the  ALEPH
experiement~\cite{Avati:2000mn},
where up to a multiplicity of $N_{\mu}\le 50$
data and simulation agree well. But as for the DELPHI
detector there is an excess of muons for higher multiplicities that cannot be
described by the simulation.
\section{Hidden Valleys and Muons}

\subsection{Basics}
Hidden sectors have been known in particle physics for quite some time. The basic idea is that besides the standard model sector there exists another hidden or dark sector in the theory that only communicates with the SM via some messenger field $X$. Usually $X$ is assumed to be quite heavy or couples only very weakly to the SM, and the non-observation of the hidden sector is attributed to this fact. This type of construction is most popular in models of supersymmetry breaking, where supersymmetry is broken in the hidden sector and communicated to the visible sector via the messenger fields. In most of these models the hidden sector is assumed to be heavy and its phenomenology is neglected. 

Only recently the possibility that the hidden sector could be relatively light has been considered. In the so-called Hidden Valley models \cite{Strassler:2006im} a new confining gauge group $G_v$ is added to the Standard Model that communicates with the SM only via a heavy $Z'$. New fermions, $v$-quarks, that are charged only under $G_v$ are confined and form $v$-hadrons. The properties of the $v$-hadrons are highly model-dependent, but the authors of \cite{Strassler:2006im} have identified some typical properties. These include long lived $v$-hadrons that decay back to SM particles with a displaced vertex and possibly also stable $v$-hadrons that provide dark matter candidates. 

Further support for Hidden Valley models with non-negligible phenomenology comes from the recent experimental results of the PAMELA experiment  that reports an excess of positrons in cosmic radiation that could originate from dark matter annihilation in the galactic halo \cite{Adriani:2008zr}. The authors of \cite{ArkaniHamed:2008qp} have shown that this can nicely be explained in terms of a hidden sector model of dark matter. In their model, dark matter is part of a dark sector with a gauge group $G_{dark}$. The corresponding dark gauge boson mixes weakly into the photon and by this mixing the sector couples to the standard model. The weak mixing allows the dark gauge boson as light as $1 \gev$ or even less, which is required to successfully explain the positron excess in terms of dark matter annihilation.

The simple Hidden Valley model \cite{Strassler:2006im} described above assumes QCD-like dynamics. Hadronization will therefore proceed similar to hadronization in QCD, so after a collision in most cases two $v$-hadron jets will form, and if the lifetimes of the $v$-hadrons are not too large, will yield events similar to those coming from a $q \bar{q} \rightarrow q\bar{q}$ scattering. However, the strongly coupled sector in the hidden valley could behave completely different from what is known from QCD. This will for example be the case if the $G_v$ gauge theory is not asymptotically free but confining in the UV instead, or, if the theory lives close to a strongly coupled fixed point. These cases have been studied in \cite{Strassler:2008bv}, and it was argued that the effects could be quite spectacular. The reason is that at large coupling (or more precicely, at large t'Hooft coupling $g_v^2 N$, where $N$ is the number of colors of the hidden gauge group $G_v$) the parton shower becomes much more efficient and thus all quarks and gluons rapidly become soft. Instead of two or three hard jets this then yields an approximately spherical distribution of soft $v$-hadrons that decay back to SM particles with an almost uniform angular distribution. 

\subsection{Hidden sector model}
For our purposes we will assume a hidden sector that has the properties described in the previous paragraph. We assume that a heavy resonance $X$ with a mass $m_X \gesim 200 \gev $ can be produced in $p \bar{p}$ collisions, either alone or accompained by a quark or a quark pair, with a quite large cross section. The resonance has a large ($\geq 99\%$) branching fraction into $v$-quark pairs that immediately shower and yield a nearly spherical distribution of $v$-hadrons. We further assume that the main decay mode of $v$-hadrons back to SM particles is into leptons, in particular muons. This could be realized for example by a mechanism similar to the one presented in \cite{ArkaniHamed:2008qp}. For definiteness we assume that the number of SM particles from one $X$ decay of the order of 100, and that about half of these particles are muons. This number will of course fluctuate largely from event to event. 

One immediate question is of course whether these events should have been seen in other experiments, in particular at colliders. The mass $m_X$ is chosen such that a production of this particle at LEP is highly unlikely, if not impossible. And it is not clear whether these events could have been detected by the LEP experiments even if $m_X$ would be much lower. The reason is that only a few if any of the produced SM particles will have energies above one $\gev$, and electromagnetically charged particles may not have been sufficiently energetic to be detected. In this context, it is interesting to note that DELPHI also observes a soft-photon excess in hadronic events. 
The same argument in principle applies to other experiments, however in section \ref{cdf} we argue that CDF might be able to see some of these muons and that they even could account for a part of the CDF muon anomaly. 
\subsection{Explaining the cosmic ray muon excess}
At collision energies above $m_X$ the $X$ resonance is copiously produced and
decays into $N$ muons and some additional SM particles. The number $N$ may
vary from event to event, but is assumed to be of order $N\sim 50$. The actual
collision energy between the partons inside the proton is on average much
lower than the center of mass energy of the proton-proton collision itself. To
obtain high enough rates, we thus require $E_{cm}\gesim 10 m_X$. For cosmic
rays consisting of protons this translates into a lower bound on the energy of
approximately $E_p \geq 10^7 \gev$, which coincides with the energy range that
accounts for most of the events with $N_\mu > 45$. For cosmic rays consisting
of iron nuclei the bound is much higher, thus we require that at least a large
fraction of cosmic rays consists of single protons or light nuclei. The
chemical composition of cosmic rays has been studied by the MACRO and EAS-TOP
collaboration\cite{Aglietta:2003hq}. They found that cosmic rays can not consist of a single particle species and that a mixture of light and heavy nuclei provides the best fit to the data, which supports our assumption. 
In its center of mass frame the $X$ resonance decays almost isotropically. The energy is distributed more or less equally between the ${\cal O} (50)$ muons, thus the energy per muon lies in the range $E_\mu \approx 1-5 \gev$. To be able to explain the excess in events with a large muon multiplicity, we must require that most of these muons hit the DELPHI detector, i.e. after travelling ${\cal O}(10^4m) $ through the atmosphere they should not deviate more than 5 meters from the track of the incoming cosmic particle. 

To be more definite, we consider the following setup: The cosmic ray propagates in the $z$-direction with 4-vector $p_{1,L} = (E_L, 0,0,E_L)$, neglecting its mass, and hits a proton at rest with mass $m_p$. We denote quantities in the laboratory (earth) frame with a subscript $L$, and quantities with respect to the center of mass of the initial collision with a subscript $CM$. The boost factor into the center of mass frame of the collision is approximately given by
\begin{align}
        \gamma_1 \approx \frac{1}{\sqrt{2}} \sqrt{\frac{E_L}{m_p}}, \qquad \beta \approx 1-\frac{m_p}{E_L} .
\end{align}
This boost factor is of order $10^3-10^4$ for the relevant energy range. For simplicity we first assume that the $X$ resonance is produced at rest. The distance $R$ between the point where the muon hits the ground and the centre of the cosmic ray shower then only depends on the energy of the muon and on the boost factor, i.e. on the energy of the incoming cosmic ray. 

\begin{figure}
  \begin{center}
  \psfig{figure=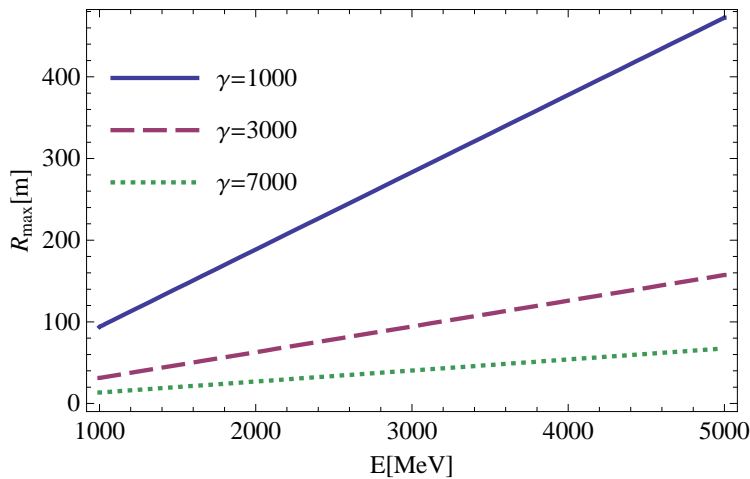}\hspace{5mm}
   \caption{Maximal distance $R_{max}$ of the muon point of impact from the shower centre, as a function of the muon energy
    in the rest frame of $X$, shown for three different boost factors and a distance $L=10 \,\text{km}$ from the surface.}
   \label{fig:rmax}
  \end{center}
\end{figure}

\begin{figure}
  \begin{center}
  \psfig{figure=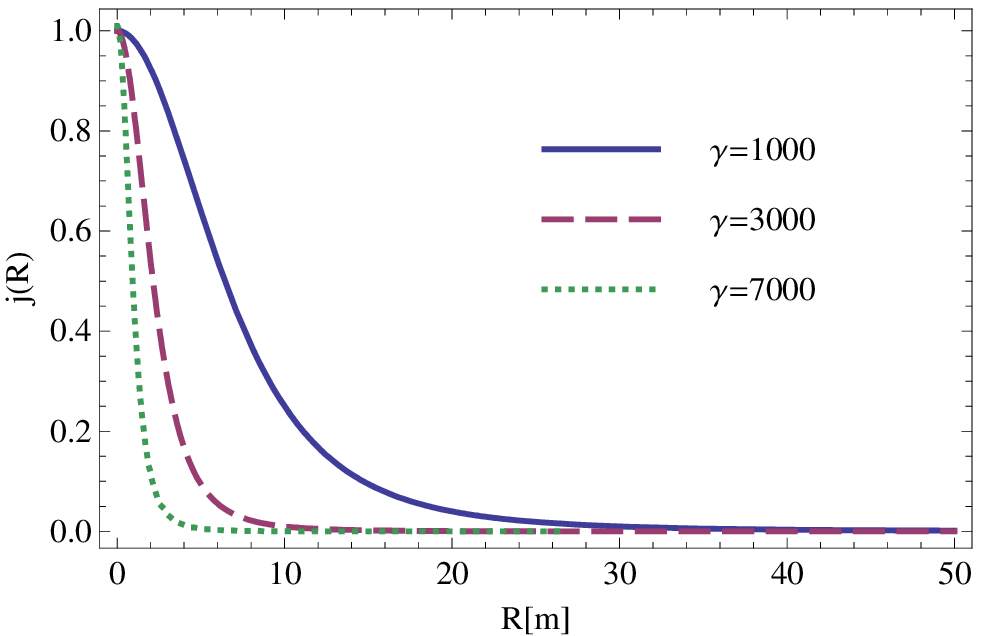, width=.46\textwidth}\hspace{5mm}
  \psfig{figure=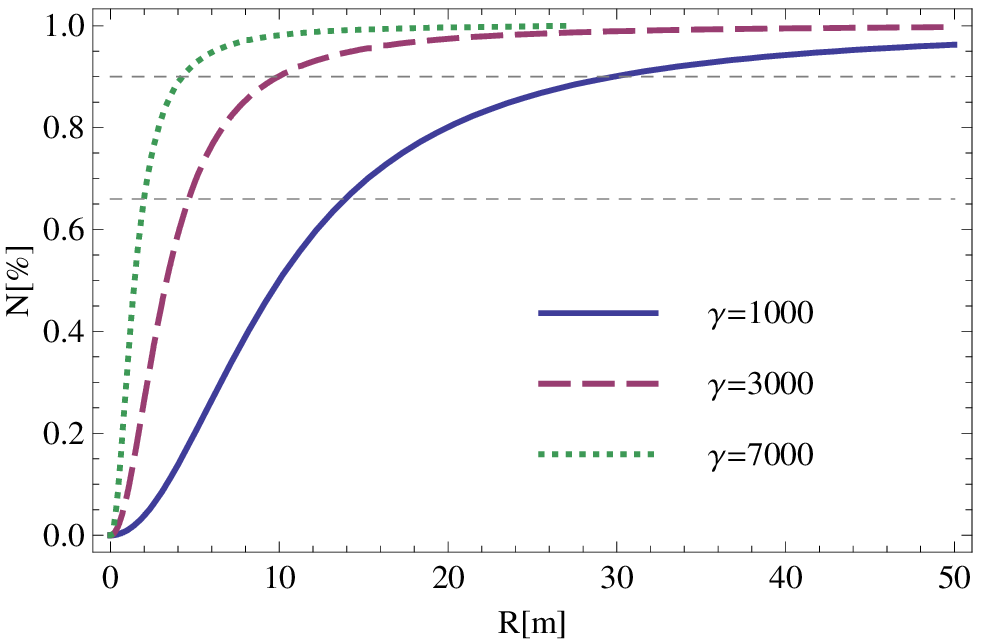, width=.46\textwidth}
  \\ a) \hspace*{7cm} b)
   \caption{The first figure shows the radial muon flux density for a muon rest frame energy $E_{\mu}=2 \gev$ for three different boost factors and $L=10\,\text{km}$, normalised to $j(0)=1$. The second figure shows the integrated distribution, i.e. the fraction of all muons that hit within a radius $R$ of the shower centre, also for $L=10\,\text{km}$.  }
   \label{fig:rd}
  \end{center}
\end{figure}

In this scenario the boost factors are large enough so that all muons hit the surface within a distance $R_{max}$ from the shower centre. The dependence of $R_{max}$ on the muon energy for different boost factors is shown in figure \ref{fig:rmax}. 
To determine the fraction of muons that actually hit the detector we need to know the radial flux density distribution $j(R)$ for a given boost factor and muon energy.
Figure \ref{fig:rd}a shows this quantity for a muon energy of $E_\mu = 2\gev$ and different boost factors. The distributions are almost identical for different muon energies. The integrated distribution, i.e. the fraction of muons that hit within a radius $R$ of the centre is shown in figure \ref{fig:rd}b. The calculation of the radial and integrated distributions is detailed in the appendix, where we also give the explicit formulas.
From figure \ref{fig:rd} we can easily see that for a boost factor $\gamma=3000$ more than two thirds of all muons hit within a radius of 5 meters from the centre, while for a boost factor larger than 7000 more than 90\% of all muons are within that range. A boost factor of $\gamma=3000$ corresponds to an energy slightly above $10^7 \gev$.
The analysis of \cite{Abdallah:2007fk} has shown that that the majority of the events with $N_\mu \geq 45$ comes from cosmic rays with an energy of $10^7-10^8 \gev$ and with the shower centre at or very close to the DELPHI detector. These are precisely the events where at least a large fraction of the muons coming from the $X$ decay would also hit the detector, since we have shown above that for this energy range most of the additional muons come within a radius of 5 meters from the centre.
We still need to estimate the fraction of events that contain an additional $X$ particle such that the excess in multi-muon events can be explained. In the DELPHI event sample
the number of events yielding 45 or more muons is around 350, the number of events with $N_\mu \geq 30$ is 1065 from table 1 in \cite{Abdallah:2007fk}. We thus assume that at least 500 events fall into the category described above, i.e. events with an initial energy $E_L> 10^7\gev$ that hit the detector almost centrally. The number of observed events with $N_\mu \geq 70$ and $N_\mu \geq 100$ can be taken from the same table, while we take the number of expected events for the same region from figure 14a in \cite{Abdallah:2007fk}. The actual numbers are shown in Tab. \ref{tab:muons}. 
\begin{table}
\begin{center}
\begin{tabular}{|l|l|l|l|}\hline 
 & experiment & proton, theory  & iron, theory \\ \hline 
$N_\mu \geq 70$  & 78 & 26 & 56 \\ \hline 
$N_\mu \geq 100$  & 24 & 6-7 & 13\\ \hline 
\end{tabular}
\caption{Observed and expected number of events at DELPHI \cite{Abdallah:2007fk} with more than 70 muons and more than 100 muons respectively.}\label{tab:muons}
\end{center}
\end{table}
Assume that a fraction $\kappa$ of the above events contains an $X$ resonance, and that in these cases on average 40 additional muons hit the detector. One can then see that with a $\kappa$ of order $5\%-10\%$ the $X$ resonance decays can account for the additional high muon-multiplicity events. 

Of course this is only a rough estimate. From the energy dependence of high energy hadron collisions, we know that $\kappa$ will not be a constant but most likely will increase for increasing cosmic ray energies. This could further enhance the number of events in the highest multiplicity range, $N_\mu >100$, which seems to be indicated by the data. In the end a full simulation of cosmic ray showers with a production of an $X$ resonance is needed to see whether this model agrees completely with the data. 
\subsection{The CDF multi-muon excess}\label{cdf}
In Ref. \cite{Aaltonen:2008qz} the CDF collaboration studied di-muon and
multi-muon events in the $b \bar{b}$ search channel. The sample consists of
events with at least two muons with an invariant mass $5 \gev \leq m_{\mu\mu} \leq 80\gev$ and transverse momentum $p_T \geq 3 \gev$. 
They further split the sample into two parts, one part where both muons are produced within the beam pipe,
and the other part where at least one muon originates from outside the beam pipe. The multiplicity of events where
both muons are produced inside the beam pipe are perfectly described by QCD production \cite{Aaltonen:2008qz}. Once the full sample is studied, approximately 150000 events remain unexplained. These events are called \textit{ghost sample} in \cite{Aaltonen:2008qz} . About half of the ghost sample can be accounted for by fake muon signals and other production mechanisms.

Additional muons
have then been searched in a cone of an angle $\theta \le 36.8^\circ$,
($\cos\theta\ge 0.8$)
around any of the initial muons in the ghost sample. It was found that the average number
of additional muons in the ghost sample are unexpectedly large, at least four times
of what one would expect for an ordinary QCD event. 

The question now is if these events 
can be explained by a hidden sector model along the lines of section 3.2.
We assume that the $X$ resonance is also produced in the $p\bar{p}$ collision at
Tevatron and decays into ${\cal O}(50)$ muons. As the hidden sector particle is heavy, of the
order of $100-200 \gev$, there is not a large boost factor between the centre-of-mass
frame and the laboratory frame. Therefore the particle will decay rather
isotropically. It is further reasonable to assume that most of these muons will have
a somewhat displaced vertex since they originate from a showering process
in the hidden sector. 

It is likely that one pair of muons from an $X$ decay has an invariant mass above the required
$5 \gev$. This determines the number of $X$ particles that must be produced in order to explain
the ghost sample to be around 70000. This number is quite large but not completely unreasonable, as has been shown for example in \cite{Domingo:2008gh}. Due to its exotic decay mode, we do not expect this process to be excluded by any existing new physics search. 

The next question then is the multiplicity of additional muons in these events. A cone with the angle of $\cos\theta\ge 0.8$ covers 10\% of the solid angle. Summing both initial muons then cover 20\%. 
We therefore expect ${\cal O}(10)$ additional muons in the cones for such a
decay. However most of these muons  will not pass the required $p_T \ge 3 \gev$, since, as discussed
above, the average energy per muon is assumed to be around $2 \gev$, but could also be much less. 
Since we do not know the energy distribution of the muons we can only give a rough estimate here. Assume that 10\% of the muons passes the $p_T$ cut. This gives us ${\cal O} (5)$ muons. Two of these will be the initial muons that make up the ghost sample. Of the remaining muons, only about 20\% will be in a cone around the initial muons. It is therefore not unlikely that the probability for a third muon is between 10\% and 30\%, and further the probability for a fourth muon is again $\sim 20\%$ of the three muon probability. Thus at least qualitatively this signal is similar to the one found by the CDF analysis. 

We do not dare to draw a final conclusion here, since too many unknowns remain. One could for example consider processes where the resonance is produced along with a recoiling particle and a sizeable boost factor. This could largely increase the number of events with very high muon multiplicities, for $M_{\chi}\ge 200 \gev$ these events would however suffer from strong kinematical suppression at Tevatron energies. We will present a more detailed study of such events at hadron colliders elsewhere. 
\section{Multi-Muon events at LHC}
\begin{figure}
  \begin{center}
  \psfig{figure=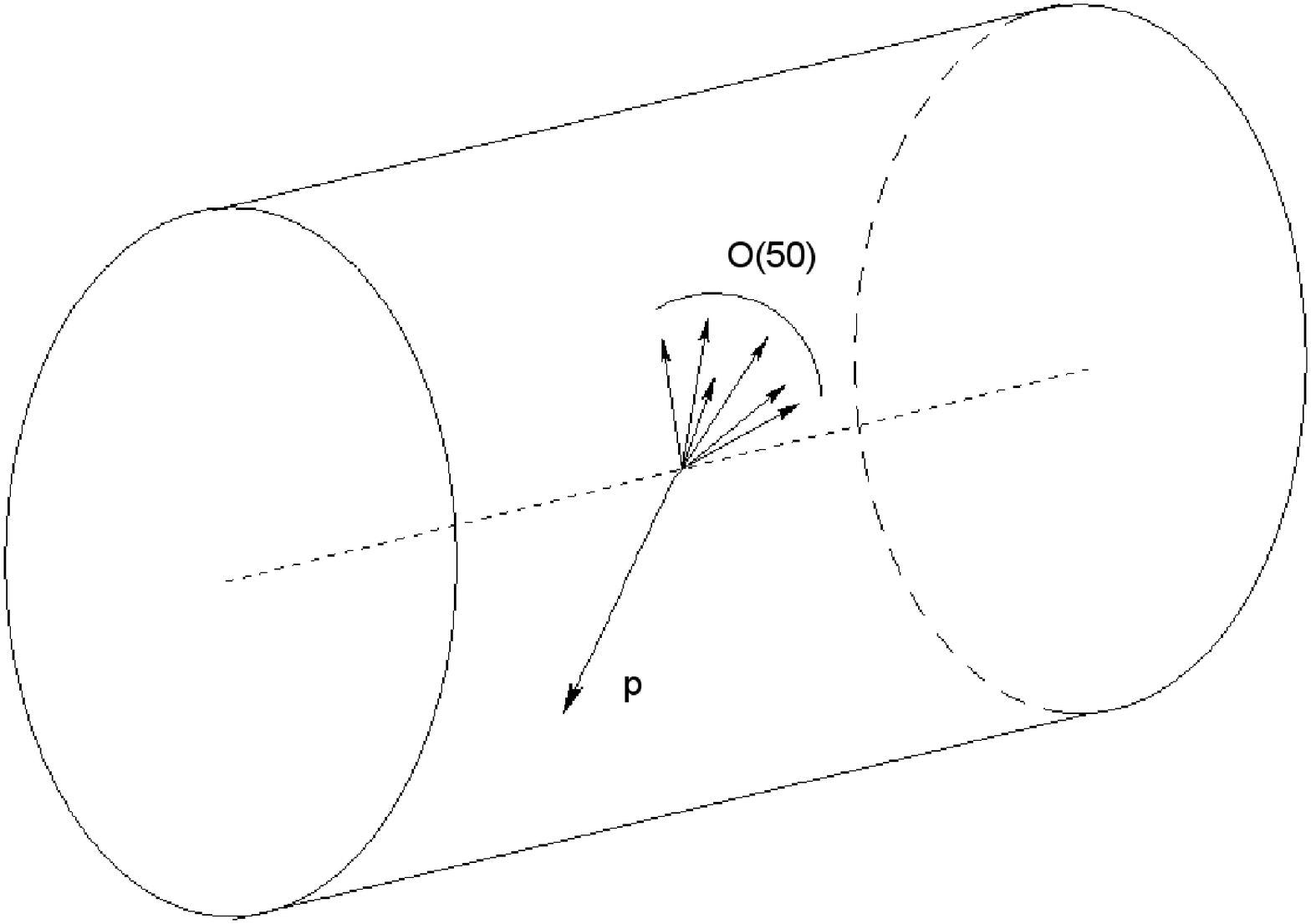, width=0.5\textwidth}
   \caption{For LHC energies the hidden sector particle can be boosted in the 
laboratory frame and therefore also leave a characteristic signature in the detector
depending on the boost factor (see figure 3).}\label{fig:lhc1}
  \end{center}
\end{figure}

For the Tevatron the centre of mass energy of $1.96 \tev$ justifies the assumption
that the hidden sector particle is produced nearly at rest and therefore
 rest frame
and laboratory frame are practically identical. This statement however no longer holds
for LHC energies. Here it is much more likely that the $X$ resonance is produced with a moderate
to large boost relative to the detector. Especially, the available collision 
energy should be sufficient to produce the $X$ resonance recoiling against 
an ordinary QCD jet.
This will lead to events as depicted in figure \ref{fig:lhc1}, where
a large number of muons with energies around $10 \gev$ are radiated in the
same direction, forming a muon jet~\cite{ArkaniHamed:2008qp}. 

\begin{figure}
  \begin{center}
  \psfig{figure=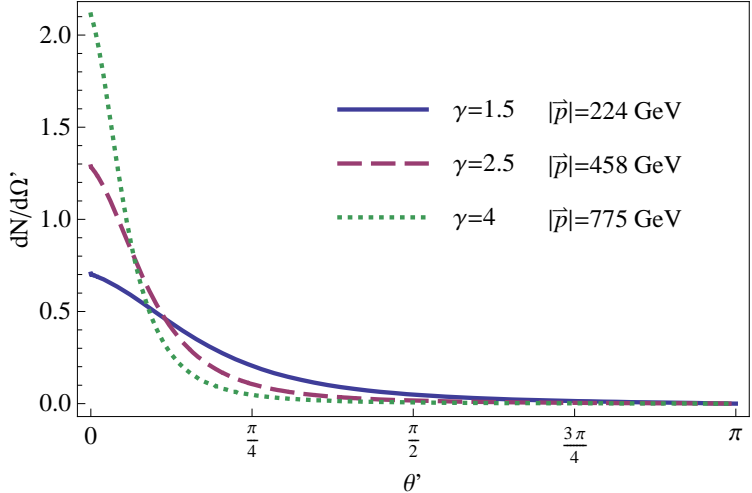}
   \caption{Angular distributions of muons in the laboratory frame, for
   different boost factors, as a function of the angle $\theta$ with respect
   to the boost direction. $\vec{p}$ is the momentum of the recoiling jet. An
   unboosted decay would produce a straight line. Distributions are normalised to unity.}\label{fig:lhc2}
  \end{center}
\end{figure}

The angular distribution of muons depends mainly on the boost factor. This
also determines the average number of muons that will hit in a cone around the
boost direction.  In figure \ref{fig:lhc2} we show the fully differential
angular distribution of the muons for different boost factors. The angle
$\theta$ always denotes the polar angle with respect to the boost direction
and with $\theta=0$ lying in direction of the boost.  For large boosts one
observes that most muons will end up in a rather narrow cone. The muon
energies are around $5-10\gev$ for smaller boost factors and up to $50\gev$
for the most energetic muons and $\gamma = 4$. The signal therefore is a large
muon jet with single particle energies between $5\gev$ and $50 \gev$ and a
small number of softer muons into other directions that might remain undetected
due to their lower energy. 

Events where the multi-muon final state is boosted 
to sufficiently high transverse momenta to be fully contained in the detector
could moreover offer the very exciting prospect of determining the 
dynamics of the $v$-parton cascade, thereby obtaining information on the
hidden gauge group. To do so, one would reconstruct the momenta of all 
produced muons, boost them back into the rest frame of $X$, and measure the 
classical event shape variables (thrust, $C$-parameter, oblateness, 
broadening) from these rest frame momenta. 
\section{Conclusion}

Hidden Valley models with conformal dynamics offer a vast and spectacular phenomenology, that 
is largely unexplored up to now. In this letter we have introduced a variant of such a model and have shown
that it can explain the excess of high multiplicity muon events in cosmic ray showers. We have further shown that the multi-muon events recently observed by the 
CDF collaboration could in principle originate from the decay of such a hidden sector resonance. 
A more rigorous analysis is required to find out whether both these signals
indeed have a common and whether a conformal hidden sector will turn out 
to be a viable explanation. Our work is the first step into this
direction. Other explanations for multi-muon production in cosmic ray events 
have been given in terms of strangelets \cite{strange} and in terms of 
anomalous multi $W$-boson production \cite{Morris:1993wg}. We have seen here 
that hidden sector models typically predict the production of soft muons in 
the CM frame of the cosmic ray interaction, and that they predict a
boost-dependent characteristic 
radial profile of the muon flux. Future measurements 
of muon production in cosmic rays 
with detectors at the surface or at shallow depth 
could provide further insight into the dynamics of these anomalous multi-muon
events. 

Alternative models have been proposed to explain the CDF anomaly
\cite{Adriani:2008zr,Giromini:2008xh}, based on cascade decays of new
particles into 8 or 10 tau leptons. While these models could turn out to be
more successful in explaining the CDF signal, they can only marginally
contribute to the muon multiplicity in cosmic ray events and therefore cannot
explain the excess seen at DELPHI and ALEPH. 

Finally we pointed out the possibly exciting signatures that such a model will
produce at the LHC. 

\bigskip

\vspace{- .3 cm}
\subsection*{Acknowledgements} We would like to thank Zoltan Kunszt for useful
discussions on the CDF multi-muon events. 
This research was supported in part by the Swiss National Science Foundation
(SNF) under contract 200020-117602 and  200020-116756/2.
\appendix

\section{Boosted Muon Densities}
Assume that the muon is emitted with an energy of $E_\mu$ in the CM frame of
the decaying $X$. 
Then its momentum is given by
\begin{align}
    p^{CM} = \big( E_\mu, p \sin \theta, 0 , p \cos \theta\big)^T, \qquad p = \sqrt{E_{\mu}^2 - m^2},
\end{align}
where $\theta$ is the angle in the CM frame under which the muon is emitted. 
In the laboratory system the momentum is given by
\begin{align}
    p^L = \big( \gamma (E_\mu -\beta p \cos\theta), p \sin \theta, 0 , \gamma(p \cos \theta - \beta E_\mu)\big)^T, \qquad p = \sqrt{E_{\mu}^2 - m^2}.
\end{align}
where $\gamma$ is the boost factor from the CM frame into the laboratory frame, and $\beta= \sqrt{1-\gamma^{-2}}$. Typical boost factors for high energy cosmic ray events range from $\gamma=10^3$ to $\gamma=10^4$. 

Assume the event takes place at a height $h$ from the surface. For large enough boost factors all muons will hit the surface. The distance of the point of impact from the shower center depends, for fixed energy, only on the angle $\theta$ under which the muon was emitted, and is given by
\begin{align}
    R (\theta) = h \frac{p_x^L(\theta)}{p_z^L(\theta)}=  h \frac{\sin\theta}{\gamma( \beta \frac{E}{p}-\cos\theta)}. \label{Rth}
\end{align}
For small boost factors or for massless particles this expression has a singularity at $\cos\theta =\beta E/p$ that determines an angle $\theta_{min}$. Muons with $\theta < \theta_{min}$ don't reach the surface but are radiated back into the universe.
For massive particles and large enough boost factors, such that $( \beta \frac{E}{p}-\cos\theta)>0$ for all $\theta$, the expression has a maximum $R_{max}$ at an angle $\theta_{max}$ that is given by
\begin{align}
    \theta_{max} = \arccos\left(\beta \frac{E_\mu}{p} \right),
\end{align}
and $R_{max}= R(\theta_{max})$.  

We now calculate the flux density as a function of the distance $R$. It is given by
\begin{align}
    j(R)=\frac{dN}{dA}\quad\text{with}\qquad dN=2\pi\sin\theta d\theta,\qquad dA=2\pi R dR.
\end{align}
Inserting $dN$ and $dA$ we obtain
\begin{align}
 j(R)=\frac{\sin\theta}{R}\frac{d\theta}{dR}.
\end{align}
For $\gamma=0$ we have $ R = h \tan\theta$ and a straightforward calculation yields
\begin{align}
 j(R)=\frac{1}{h^2}\cos^3\theta(R).
\end{align}

For $\gamma \neq 0$ the function $d\theta/dR$ can be determined from equation (\ref{Rth}). The final expression is
\begin{align}
        j(R)  = \frac{\sin\theta}{R}\times \frac{\gamma}{h} \left( \frac{\cos\theta}{(\beta \frac{E}{p}-\cos\theta)} - \frac{\sin^2\theta}{(\beta\frac{E}{p}-\cos\theta)^2}\right)^{-1},
\end{align}
where $\theta = \theta(R)$ has to be determined from the inverse of eqn. (\ref{Rth}). The function $R(\theta)$ is not monotonic over the full range of $\theta$. One has to distinct the two cases $\theta <\theta_{max}$ and $\theta > \theta_{max}$ and add both contributions to obtain the full density profile. As it turns out for large boost factors the contribution from $\theta < \theta_{max}$ can be neglected. 

\end{document}